\documentclass[twocolumn,aps,superscriptaddress]{revtex4}

\usepackage{amssymb}
\usepackage{amsmath}
\usepackage{graphicx}
\usepackage[normalem]{ulem}
\usepackage[dvips]{color}
\usepackage{appendix}

\begin{document}

\title{Bayesian inference of nuclear symmetry energy from measured and imagined neutron skin thickness in $^{116,118,120,122,124,130,132}$Sn, $^{208}$Pb, and $^{48}$Ca}

\author{Jun Xu\footnote{xujun@zjlab.org.cn}}
\affiliation{Shanghai Advanced Research Institute, Chinese Academy of Sciences,
Shanghai 201210, China}
\affiliation{Shanghai Institute of Applied Physics, Chinese Academy
of Sciences, Shanghai 201800, China}
\author{Wen-Jie Xie\footnote{wenjiexie@yeah.net}}
\affiliation{Department of Physics, Yuncheng University, Yuncheng 044000, China}
\author{Bao-An Li\footnote{Bao-An.Li@Tamuc.edu}}
\affiliation{Department of Physics and Astronomy, Texas A$\&$M University-Commerce, Commerce, TX 75429, USA}

\date{\today}

\begin{abstract}
The neutron skin thickness $\Delta r_{np}$ in heavy nuclei has been known as one of the most sensitive terrestrial probes of the nuclear symmetry energy $E_{\rm{sym}}(\rho)$ around $\frac{2}{3}$ of the saturation density $\rho_0$ of nuclear matter. Existing neutron skin data mostly from hadronic observables suffer from large uncertainties and their extraction from experiments are often strongly model dependent. While waiting eagerly for the promised model-independent and high-precision neutron skin data for $^{208}$Pb and $^{48}$Ca from the parity-violating electron scattering experiments (PREX-II and CREX at JLab as well as MREX at MESA), within the Bayesian statistical framework using the Skyrme-Hartree-Fock model we infer the posterior probability distribution functions (PDFs) of the slope parameter $L$ of the nuclear symmetry energy at $\rho_0$ from imagined $\Delta r_{np}(^{208}$Pb$)=0.15$, 0.20, and 0.30 fm with a $1\sigma$ error bar of 0.02, 0.04, and 0.06 fm, respectively, as well as $\Delta r_{np}(^{48}$Ca$)=0.12$, 0.15, and 0.25 fm with a $1\sigma$ error bar of 0.01 and 0.02 fm, respectively. The results are compared with the PDFs of $L$ inferred using the same approach from the available $\Delta r_{np}$ data for $^{116,118,120,122,124,130,132}$Sn from hadronic probes. They are also compared with results from a recent Bayesian analysis of the radius and tidal deformability data of canonical neutron stars from GW170817 and NICER. The neutron skin data for Sn isotopes gives $L=45.5^{+26.5}_{-21.6}$ MeV surrounding its mean value or $L=53.4^{+18.6}_{-29.5}$ MeV surrounding its maximum {\it a posteriori} value, respectively, with the latter smaller than but consistent with the $L=66_{-20}^{+12}$ MeV from the neutron star data within their 68\% confidence intervals. We found that $\Delta r_{np}=0.17 - 0.18$ fm in $^{208}$Pb with an error bar of about 0.02 fm leads to a PDF of $L$ compatible with that from analyzing the Sn data. In order to provide additionally useful information on $L$ extracted from the $\Delta r_{np}$ of Sn isotopes, the experimental error bar of $\Delta r_{np}$ in $^{208}$Pb should be at least smaller than 0.06 fm aimed by some current experiments. In addition, the $\Delta r_{np}(^{48}$Ca) needs to be larger than 0.15 fm but smaller than 0.25 fm to be compatible with the Sn and/or neutron star results. To further improve our current knowledge about $L$ and distinguish its PDFs in the examples considered, even higher precisions leading to significantly less than $\pm 20$ MeV error bars for $L$ at 68\% confidence level are necessary.

\end{abstract}

\maketitle

\section{INTRODUCTION}
\label{introduction}

Nuclear symmetry energy $E_{sym}(\rho)$ encodes the information about the energy necessary to make nuclear systems, such as nuclei, neutron stars, and matter created during collisions of two nuclei or neutron stars, more neutron rich \cite{Tesym}. As such, reliable knowledge about the symmetry energy has broad impacts on many critical issues in both nuclear physics and astrophysics \cite{Ste05,Lat07,Li19}. Thanks to the great efforts of many people in both communities over the last two decades, much progress has been made in constraining both the magnitude $E_{sym}(\rho_0)$ and the slope parameter $L=3\rho_0(dE_{sym}/d\rho)_{\rho_0}$ at the saturation density $\rho_0$ of nuclear matter \cite{Bar05,Li08,Tsa12,Tra12,Hor14,Bal16,Li17,PPNP-Li}. For example, fiducial values of $E_{\rm{sym}}(\rho_0)=31.7 \pm 3.2$~MeV and $L = 58.7 \pm 28.1$~MeV were found from surveying 53 analyses \cite{BAL13,Oer17} carried out by 2016 using various terrestrial nuclear laboratory data and astrophysical observations. In comparison, using a novel Bayesian approach to quantify the truncation errors in chiral effective field theory (EFT) predictions for pure neutron matter and a many-body perturbation theory with consistent nucleon-nucleon and three-nucleon interactions up to fourth order in the EFT expansion, the $E_{\rm{sym}}(\rho_0)$ and $L$ were found very recently to be $E_{\rm{sym}}(\rho_0)=31.7 \pm 1.1$~MeV and $L = 59.8 \pm 4.1$~MeV \cite{Ohio20}, respectively. In a very recent Bayesian analysis of the radius and tidal deformability data of canonical neutron stars from GW170817 and NICER, the most probable value of $L=66_{-20}^{+12}$ MeV at 68\% confidence level was found \cite{Xie19} while the $E_{sym}(\rho_0)$ remains the same as the fiducial value. Clearly, these results are all highly consistent while the error bars from the data analyses are significantly larger than the EFT predictions. One of the possible reasons for the larger error bars of the extracted $L$ values is that the extraction of the symmetry energy from terrestrial experiments often involves large and sometimes unqualifiable theoretical uncertainties. Moreover, exiting experimental data are mostly from hadronic probes that are known to suffer from large statistical and systematical errors. Thus, in the continuous strive to better constrain the density dependence of nuclear symmetry energy, significant efforts are being made in the nuclear physics community to better quantify theoretical uncertainties and/or to find more clean experimental probes, see, e.g., Refs.~\cite{Naz14,Jorge19}.

The neutron skin thickness $\Delta r_{np}=R_n-R_p$ is the difference in root-mean-square neutron $R_n$ and proton $R_p$ radii. The $\Delta r_{np}$ values of heavy nuclei have been known as one of the most sensitive terrestrial probes of the nuclear symmetry energy $E_{\rm{sym}}(\rho)$ at subsaturation densities around $\frac{2}{3}\rho_0$, see, e.g., Refs.~\cite{Bro00,Typ01,Chuck01,Fur02,Tod05,Cen09,Zha13,India1,India2}. For recent reviews, we refer the readers to Refs. \cite{Thiel,Burg}. It has been shown using various nuclear many-body theories that the $\Delta r_{np}$ is approximately proportional to the density slope within theoretical uncertainties, see, e.g., Refs. \cite{Vin14,X18} for reviews. In fact, considerable efforts have been devoted continuously to measuring the $\Delta r_{np}$ in $^{208}$Pb for decades \cite{Ray78}. For earlier reviews, see, e.g., Refs. \cite{Ray-PR,Sta94}. More recently, for example, $\Delta r_{np}=0.211^{+0.054}_{-0.063}$ fm and $0.16\pm0.07$ fm were obtained from proton~\cite{Zen10} and pion~\cite{Fri12} scatterings, respectively. Studies from the annihilation of antiprotons on the nuclear surface gave $\Delta r_{np}=0.18\pm 0.04({\rm expt.}) \pm 0.05({\rm theor.})$ fm~\cite{Bro07,Klo07}, while the isospin diffusion data in heavy-ion collisions imply $\Delta r_{np}$ to be around $0.22\pm0.04$ fm~\cite{SteLi05,Che05}, and $\Delta r_{np}=0.15 \pm 0.03 ({\rm stat.}) ^{+0.01}_{-0.03} ({\rm sys.})$ fm was obtained from coherent pion photoproductions~\cite{Tar14}. Obviously, both the mean and error bar of $\Delta r_{np}(^{208}$Pb) are not well determined. Consequently, in studying impacts of $\Delta r_{np}(^{208}$Pb) on neutron stars, sometimes a fiducial value of $\Delta r_{np}(^{208}$Pb$)=0.20\pm 0.04$ fm was used \cite{Ste05,LiSte}. Among the available data for heavy nuclei, the $\Delta r_{np}$ of Sn isotopes have been most extensively measured using isovector spin-dipole resonances excited by the charge-exchange reactions~\cite{Kra99}, antiproton annihilations~\cite{Trz01}, and proton elastic scatterings~\cite{Ter08}, etc. We will therefore first use the measured $\Delta r_{np}$ values of Sn isotopes to establish a reference PDF for $L$ in our Bayesian analyses, and compare the results with the information from a traditional approach using forward-modeling with $\chi^2$ minimization. This reference serves as a quantitative measure of our current knowledge about inferring the $L$ value using available neutron skin data. We will then measure possible improvements to this knowledge by using anticipated high-precision neutron skin data for $\Delta r_{np}(^{208}$Pb) and $\Delta r_{np}(^{48}$Ca) from parity-violating electron-nucleus scattering experiments.

While most of the available neutron skin data from hadronic probes suffer from large statistical and systematic errors as well as model dependence, parity-violating electron scatterings were shown theoretically to provide model-independent and high-precision measures of neutron skin thickness \cite{Chuck1,Chuck2}. However, these experiments are extremely difficult. While the pioneering Lead ($^{208}$Pb) Radius EXperiment (PREX) at the Jeferson Laboratory (JLab), i.e., PREX-I experiment, has demonstrated an excellent control of systematic errors, the resulting $\Delta r_{np}(^{208}$Pb$)=0.33^{+0.16}_{-0.18}$ fm still has a large error bar~\cite{Abr12}. The PREX-II experiment and the Calcium Radius EXperiment (CREX) at Jlab are expected to dramatically reduce the error bars to the level of $\pm 0.06$ fm for $^{208}$Pb and $\pm 0.02$ fm for $^{48}$Ca, respectively~\cite{Thiel}. Even better, the planned Mainz Radius EXperiment (MREX) at the Mainz Energy recovery Superconducting Accelerator (MESA) aims to determine the neutron radius in $^{208}$Pb with a 0.5\% (or 0.03 fm) precision; while for $^{48}$Ca the sensitivity is similar to the one expected from the CREX at JLab \cite{Thiel}. If realized, these experiments may improve dramatically our knowledge about the nuclear symmetry energy and help constrain tightly nuclear theories.

Wishing the experimentalists all the best luck in the world and eagerly waiting for their new results from the parity-violating electron scattering experiments, hinted by existing results and the planned experiments, we imagine a few mean values and error bars for the neutron skin thickness in $^{208}$Pb and $^{48}$Ca in our Bayesian inference of the symmetry energy slope parameter $L$. We compare the resulting PDFs of $L$ with those from Bayesian analyses of neutron star observations and the neutron skin thickness in Sn isotopes. Following the spirit of a recent work conducting covariant analysis to obtain analytic insights on the information content of new observables \cite{Jorge20}, we also try to answer the two questions posted by Reinhard and Nazarewicz \cite{Rei10}: (1) {\it Considering the current theoretical knowledge, what novel information does new measurements bring in?} and (2) {\it How can new data reduce the uncertainties of current theoretical models?} More specifically, we study (1) {\it How the uncertainties of the neutron skin measurements affect the extraction of the symmetry energy?} and (2) {\it What additional information about the symmetry energy can new measurements bring to us?} In order to address these questions, besides comparing with results from Bayesian analyses of the very recent data from neutron star observations and the old neutron skin data of Sn isotopes, we freely dreamed that the experimentalists would some day measure the $\Delta r_{np}(^{208}$Pb) and $\Delta r_{np}(^{48}$Ca) at precisions even better than they already planned at Jlab and/or MESA. We understand that these will be extremely challenging, but we assume that they are not more difficult than measuring nuclear matter effects on the strain amplitude and frequency of gravitational waves from merging neutron stars.

The rest of the paper is arranged as follows. We shall first summarize in Section \ref{theory} the most relevant aspects of the standard Skyrme-Hartree-Fock (SHF) model and interactions we use, and then recall the main formalisms and prior information we use in the Bayesian analyses. In Section \ref{results} we present and discuss our results. The summary and conclusions are given in Section \ref{summary}.

\section{Theoretical framework}
\label{theory}

Within the Bayesian statistical framework we infer from the neutron skin thickness data the posterior PDFs of isovector nuclear interactions used in the standard SHF model. These isovector interactions determine the density dependence of nuclear symmetry energy, while the isoscalar parameters are fixed at their currently known most probable values. Consequently, the posterior PDF of the symmetry energy slope parameter $L$ can be obtained. For completeness and ease of discussions, we summarize in the following the most important aspects of the SHF model and the Bayesian approach as well as the specific inputs used in this work. We skip most of the details that can be found easily in the literature.

\subsection{Skyrme-Hartree-Fock model}
\label{shf}

We start from the following standard effective Skyrme interaction between nucleon 1 and nucleon 2~\cite{Cha97}
\begin{eqnarray}\label{SHFv}
v(\vec{r}_1,\vec{r}_2) &=& t_0(1+x_0P_\sigma)\delta(\vec{r}) \notag \\
&+& \frac{1}{2} t_1(1+x_1P_\sigma)[{\vec{k}'^2}\delta(\vec{r})+\delta(\vec{r})\vec{k}^2] \notag\\
&+&t_2(1+x_2P_\sigma)\vec{k}' \cdot \delta(\vec{r})\vec{k} \notag\\
&+&\frac{1}{6}t_3(1+x_3P_\sigma)\rho^\alpha(\vec{R})\delta(\vec{r}) \notag\\
&+& i W_0(\vec{\sigma}_1+\vec{\sigma_2})[\vec{k}' \times \delta(\vec{r})\vec{k}].
\end{eqnarray}
In the above, $\vec{r}=\vec{r}_1-\vec{r}_2$ and $\vec{R}=(\vec{r}_1+\vec{r}_2)/2$ are related to the positions of two nucleons $\vec{r}_1$ and $\vec{r}_2$, $\vec{k}=(\nabla_1-\nabla_2)/2i$ is the relative momentum operator and $\vec{k}'$ is its complex conjugate acting on the left, and $P_\sigma=(1+\vec{\sigma}_1 \cdot \vec{\sigma}_2)/2$ is the spin exchange operator, with $\vec{\sigma}_{1(2)}$ being the Pauli matrics. The parameters $t_0$, $t_1$, $t_2$, $t_3$, $x_0$, $x_1$, $x_2$, $x_3$, and $\alpha$ determine macroscopic quantities describing the saturation properties of symmetric nuclear matter, density dependence of nuclear symmetry energy, and structures of finite nuclei. Inversely, they can be expressed analytically in terms of several macroscopic quantities, facilitating the Bayesian inference of the latter directly from the neutron skin data. In this work, we use the MSL0 interaction~\cite{MSL0}.
Specifically,  the macroscopic quantities used are: the saturation density $\rho_0$, the binding energy $E_0$ at the saturation density, the incompressibility $K_0$, the isoscalar and isovector nucleon effective mass $m_s^\star$ and $m_v^\star$ at the Fermi momentum in normal nuclear matter, the symmetry energy $E_{sym}^0\equiv E_{\rm{sym}}(\rho_0)$ and its slope parameter $L$ at the saturation density, and the isoscalar and isovector density gradient coefficient $G_S$ and $G_V$. The spin-orbit coupling constant is fixed at $W_0=133.3$ MeVfm$^5$. In the present study, we calculate the posterior PDFs of the isovector interaction parameters, i.e., $E_{sym}^0$, $L$, and $m_v^\star$, by varying them randomly with equal probability within their respective prior ranges, while fixing the other macroscopic quantities at their empirical values as in the original MSL0 interaction~\cite{MSL0}.

The potential energy density can be calculated from the above effective interaction [Eq.~(\ref{SHFv})] based on the Hartree-Fock method, and the single-particle Hamitonian can then be obtained using the variational principle. Here we assume that the nucleus is spherical and only time-even contributions are considered. Solving the Schr\"odinger equation leads to the wave functions of each nucleon, and the density distributions for neutrons and protons can be calculated accordingly. The neutron skin thickness can then be obtained from the difference of the root-mean-square radii between neutrons and protons. For details of this standard procedure, we refer the reader to Ref.~\cite{Vau72}. In the present work, we use Reinhard's SHF code described in Ref.~\cite{SHFcode}.

\subsection{Bayesian analysis}
\label{bayes}

Compared to the traditional approach of forward-modeling together with a $\chi^2$ minimization to fit the experimental data and empirical properties of nuclear matter, the advantages of Bayesian analysis in the uncertainty quantification and evaluating correlations of model parameters have been well documented in the literature, see, e.g. Ref. \cite{Kej20} for a very recent overview of the Bayesian approach and its applications in studying nuclear structures. We adopt it here to infer the posterior PDFs of the isovector interaction parameters and the corresponding nuclear symmetry energy from the neutron skin data. The Bayes' theorem describes how new experimental data may improve a hypothesis reflecting prior knowledge via
\begin{equation}
P(M|D) = \frac{P(D|M)P(M)}{\int P(D|M)P(M)dM}.
\end{equation}
In the above, $P(M|D)$ is the posterior PDF for the model $M$ given the data set $D$, $P(D|M)$ is the likelihood function or the conditional probability for a given theoretical model $M$ to predict correctly the data $D$, and $P(M)$ denotes the prior PDF of the model $M$ before being confronted with the data. The denominator of the right-hand side of the above equation is the normalization constant.

For the prior PDFs, we choose the model parameters $p_1=E_{sym}^0$ uniformly within $25 \sim 35$ MeV, $p_2=L$ uniformly within $0 \sim 120$ MeV, and $p_3=m_v^\star/m$ uniformly within $0.5 \sim 1$, with $m$ being the bare nucleon mass. Our choice of the large prior range and the uniform PDF for the symmetry energy slope parameter $L$ is intentionally ignorant with respect to our current knowledge from many earlier analyses of both terrestrial and astrophysical data as well as the state-of-the-art EFT predictions as we outlined in the introduction. Without belittling the invaluable prior knowledge from the hard work of many people over two decades, this choice helps us reveal how the neutron skin data alone may narrow down the uniform prior PDF of $L$ in the artificially enlarged range of $0\sim120$ MeV.

For a given set of the MSL0 interaction parameters, the theoretical neutron skin thickness $d^{th}_1=\Delta r_{np}^{(1)}$, $d^{th}_2=\Delta r_{np}^{(2)}$, ... for different nuclei from the SHF calculations are used to calculate the likelihood of these model parameters with respect to the corresponding experimental data $d^{exp}_1$, $d^{exp}_2$, ... according to
\begin{eqnarray}
&&P[D(d_{1,2,...})|M(p_{1,2,3})] \notag\\
&=& \Pi_{i} \frac{1}{\sqrt{2\pi} \sigma_i}\exp\left[-\frac{(\Delta d_i)^2}{2\sigma_i^2}\right], \label{llh}
\end{eqnarray}
where $\Delta d_i$ and $\sigma_{i}$ denote respectively the deviation of theoretical results from the experimental data and the width in the likelihood function from an independent experimental data sample $i$. In principle, the likelihood function depends on uncertainties of both the experimental data and model predictions. For the neutron-skin thickness of $^{208}$Pb or $^{48}$Ca, we use the imagined experimental error bar (which is varied and could be considered as due to both experimental and model uncertainties) as the width $\sigma_i$ as often done in the literature, and $\Delta d_i = |d^{th}_i-d^{exp}_i|$ being the deviation of the theoretical result from the mean value of the imagined experimental data. For the neutron skin thicknesses of $^{116,118,120,122,124}$Sn~\cite{Ter08} and $^{130,132}$Sn~\cite{Kli07}, they are treated as two independent experimental data samples with each series extracted from a correlated method. The deviation and the width for each series are calculated according to
\begin{eqnarray}
\Delta d_i &=& \sqrt{\sum_j^{j \in i} (d_j^{th}-d_j^{exp})^2}, \label{di}\\
\sigma_i &=& \sqrt{\sum_j^{j \in i} \sigma_j^2}, \label{si}
\end{eqnarray}
where $d_j^{th}$, $d_j^{exp}$, and $\sigma_j$ are the theoretical result, the mean value of the experimental data, and the experimental $1\sigma$ error bar of the neutron skin thickness for a Sn isotope $j$, respectively.

The posterior PDF of a single model parameter $p_i$ is given by
\begin{equation}\label{1dpdf}
P(p_i|D) = \frac{\int P(D|M) P(M) \Pi_{j\ne i} dp_j}{\int P(D|M) P(M) \Pi_{j} dp_j},
\end{equation}
while the correlated PDF of two model parameters $p_i$ and $p_j$ is given by
\begin{equation}\label{2dpdf}
P[(p_i,p_j)|D] = \frac{\int P(D|M) P(M) \Pi_{k\ne i,j} dp_{k}}{\int P(D|M) P(M) \Pi_{k} dp_k}.
\end{equation}
For the one-dimensional PDF, the range of the model parameter at the $68\%$ confidence level is obtained according to~\cite{Tur93}
\begin{equation}
\int_{p_{iL}}^{p_{iU}}  P(p_i|D) dp_i = 0.68, \label{confidence}
\end{equation}
where $p_{iL}$ ($p_{iU}$) is the lower (upper) limit of the corresponding narrowest interval of the parameter $p_i$ surrounding its mean value
\begin{equation}
\langle p_i \rangle = \int p_i P(p_i|D) dp_i \label{mean}
\end{equation}
or its maximum {\it a posteriori} (MAP) value. The calculation of the posterior PDFs is based on the Markov-Chain Monte Carlo approach using the Metropolis-Hastings algorithm~\cite{Met53,Has70}. The calculation generally takes about $10^5-10^6$ steps, and the analysis is carried out after the first $10^4$ steps when the convergence is mostly reached.

\section{Results and discussions}
\label{results}

\subsection{$L$ from measured neutron skin thickness in $^{116,118,120,122,124,130,132}$Sn}

To establish a reference for comparisons, we first perform Bayesian analyses with the real experimental data of neutron skin thicknesses $\Delta r_{np}$ in Sn isotopes. It is one of the most complete $\Delta r_{np}$ data sets along the longest isotope chain available. The $\Delta r_{np}$ data of $^{116,118,120,122,124}$Sn and $^{130,132}$Sn as two independent experimental data samples from Refs.~\cite{Ter08,Kli07} are listed in Table~\ref{T1}. The mean values and the experimental $1\sigma$ errors together with the theoretical results are used in calculating the likelihood function according to Eqs.~(\ref{llh}), (\ref{di}), and (\ref{si}).

\begin{figure*}[ht]
\includegraphics[scale=0.22]{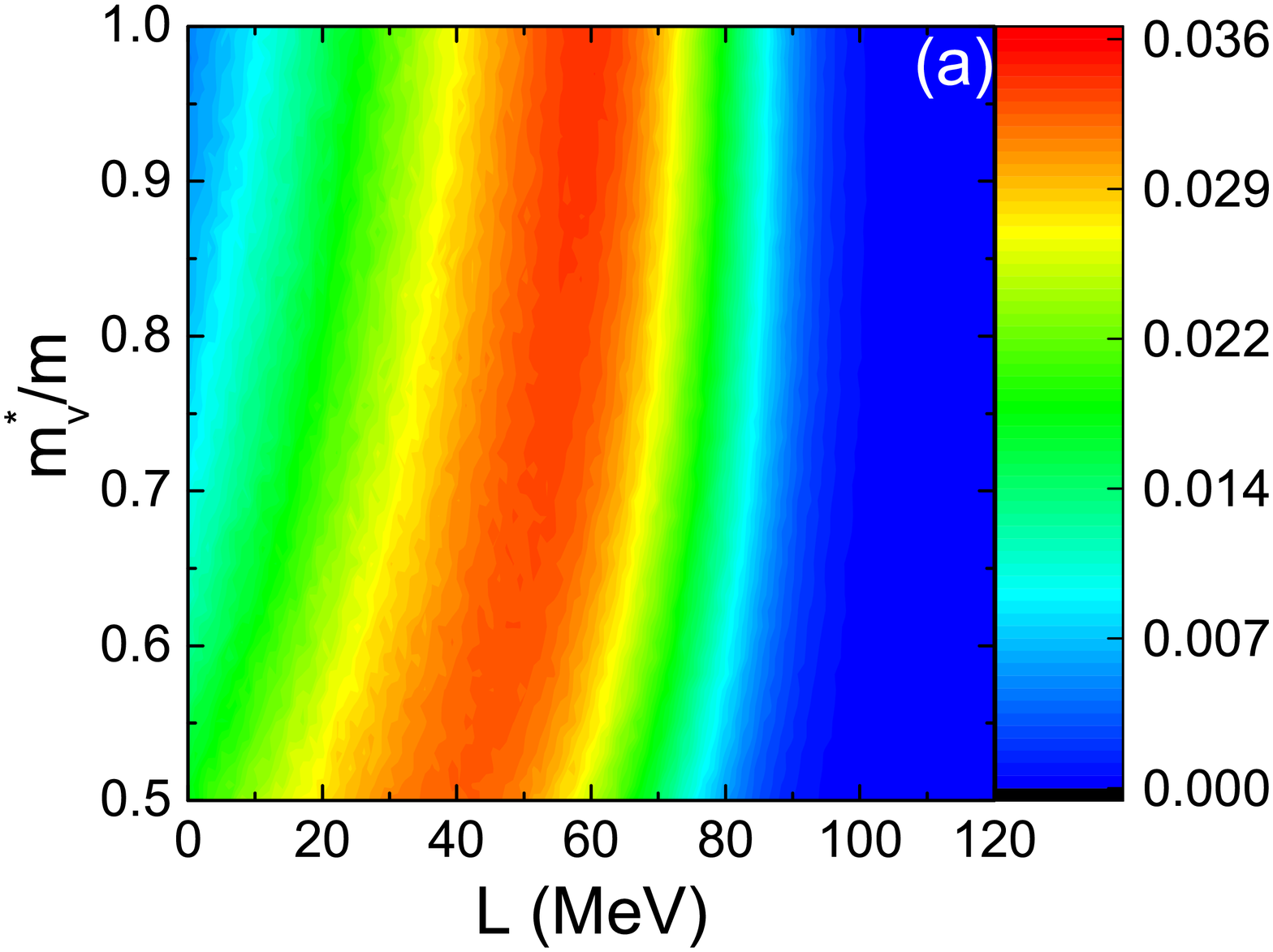}
\includegraphics[scale=0.22]{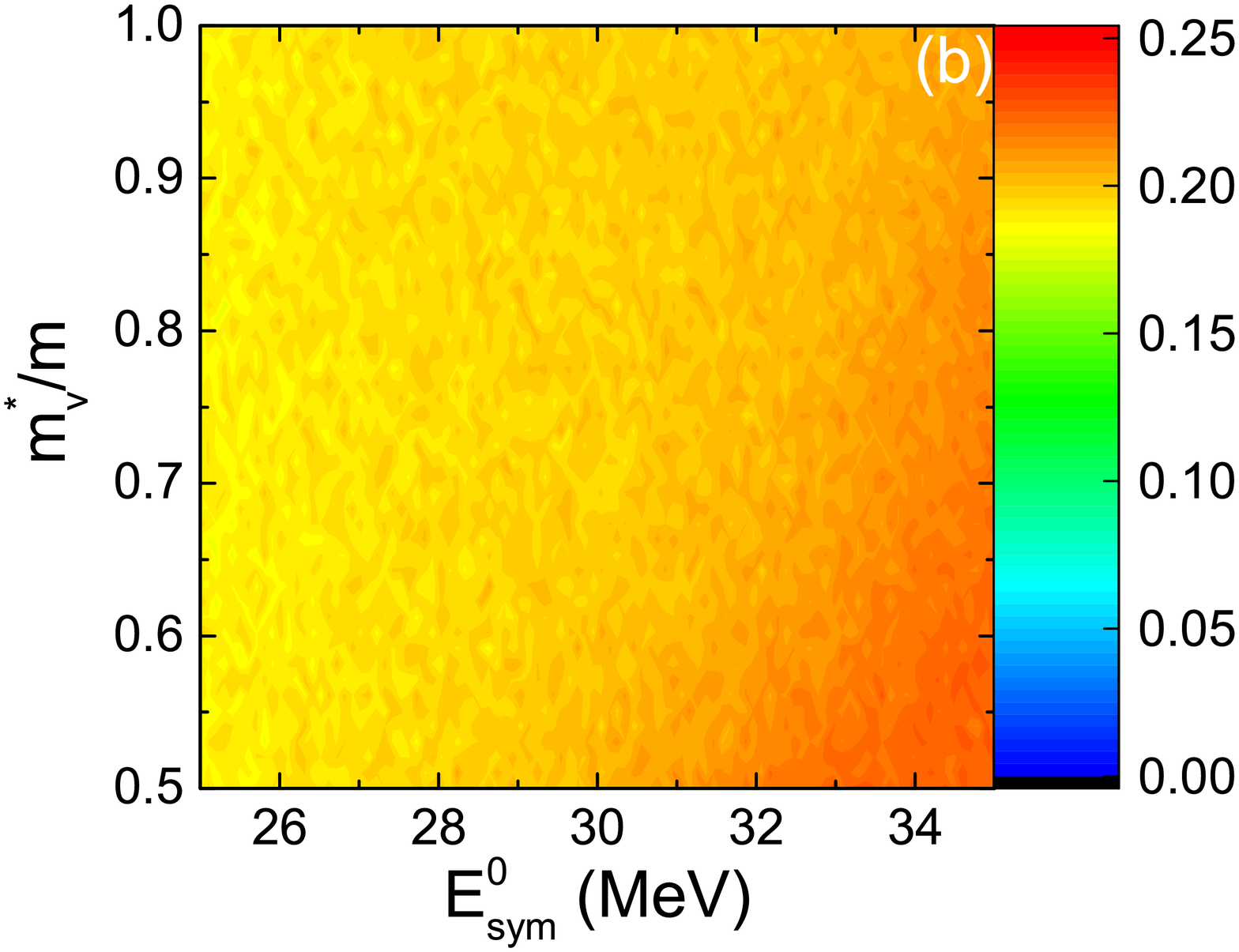}
\includegraphics[scale=0.22]{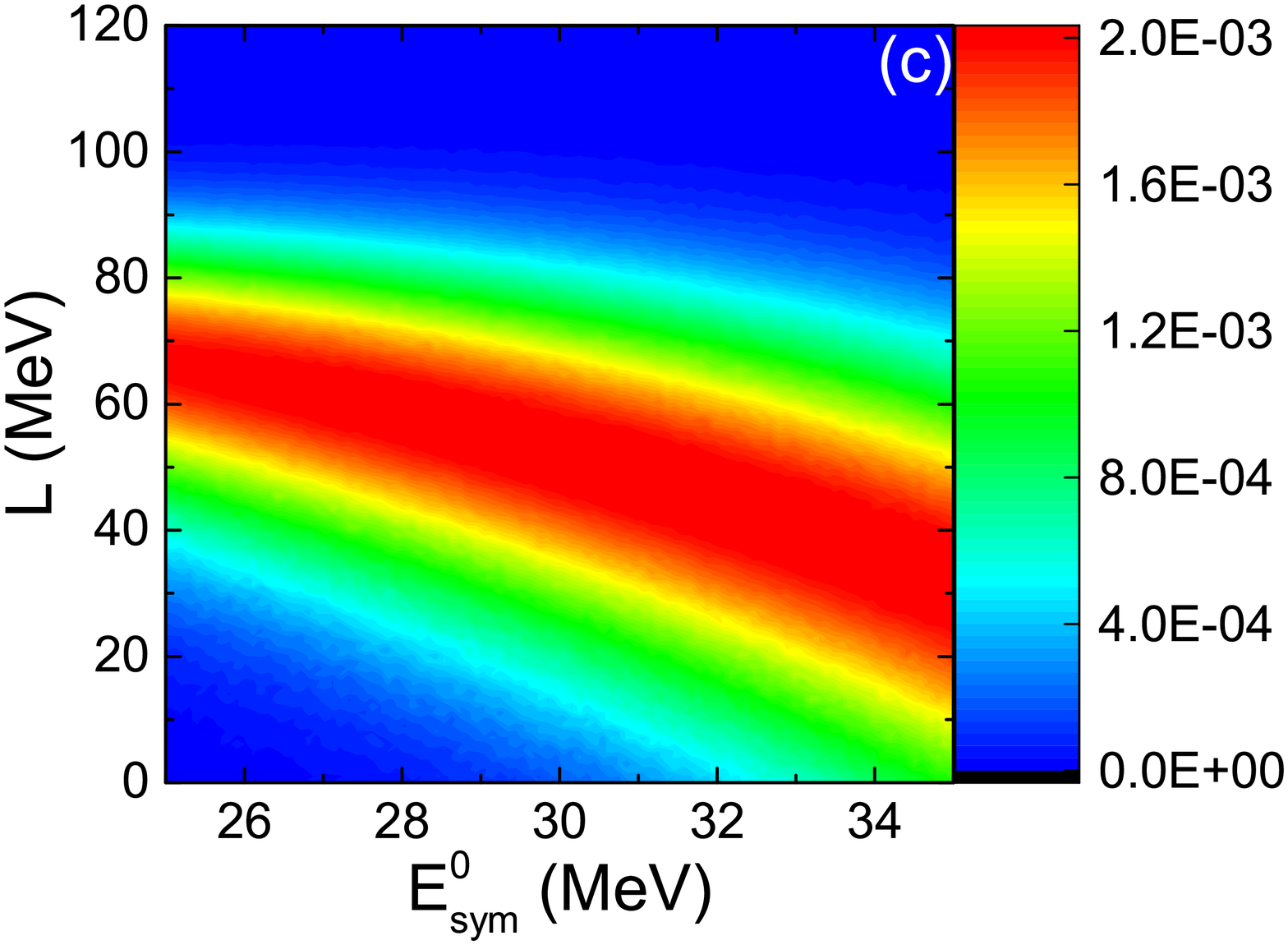}\\
\vspace{0.4cm}
\includegraphics[scale=0.24]{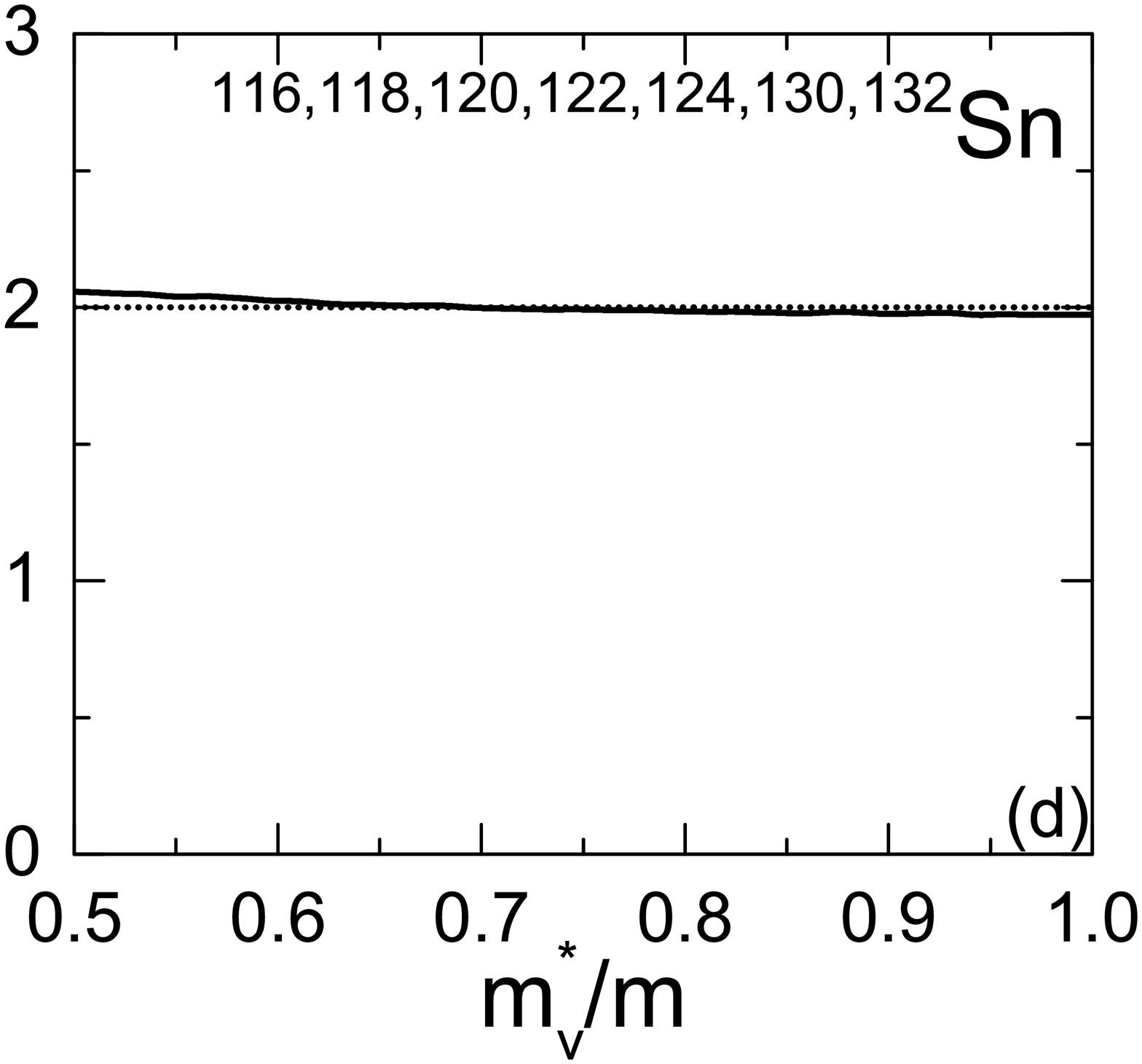}
\hspace{0.3cm}\includegraphics[scale=0.24]{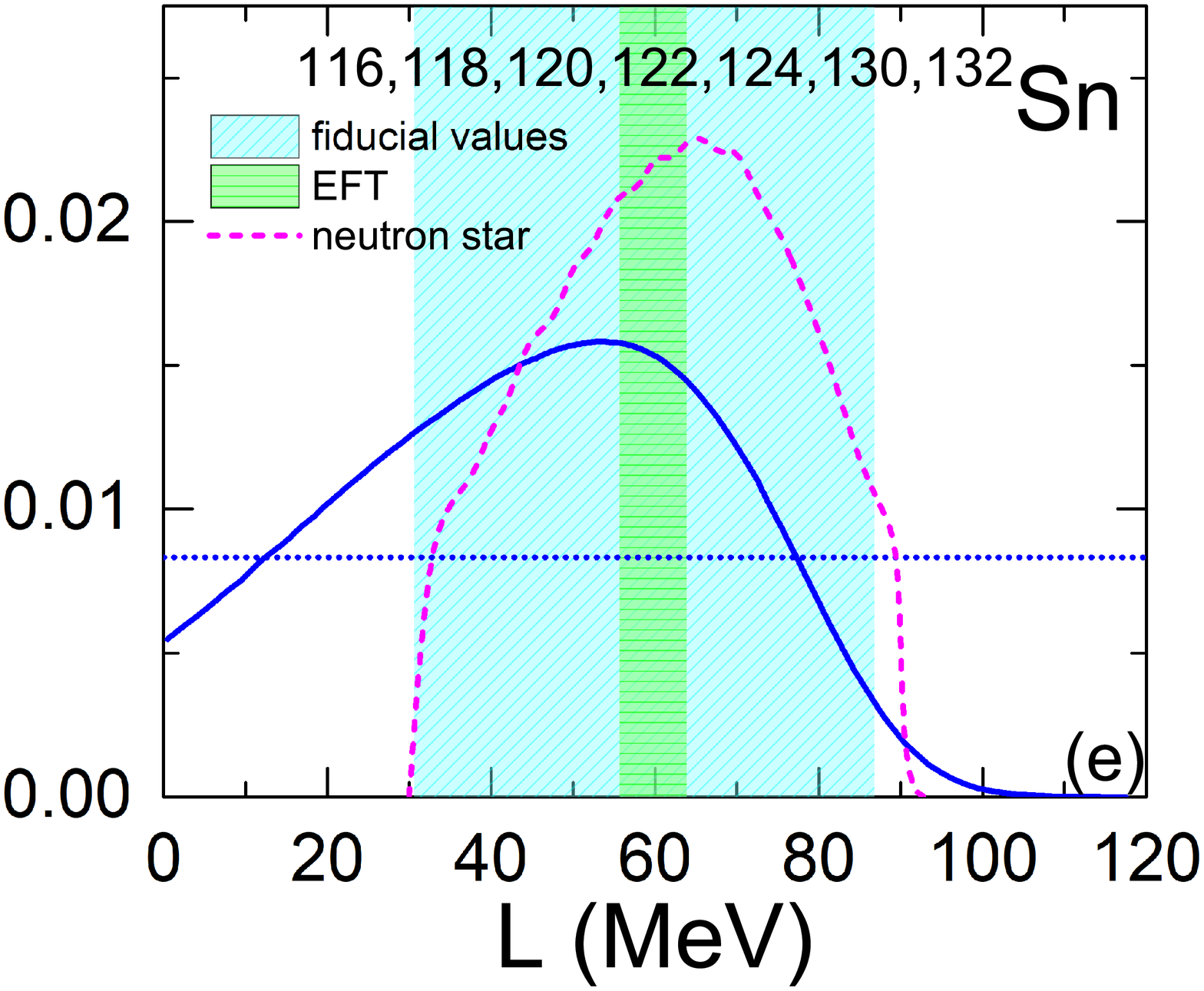}
\hspace{0.3cm}\includegraphics[scale=0.24]{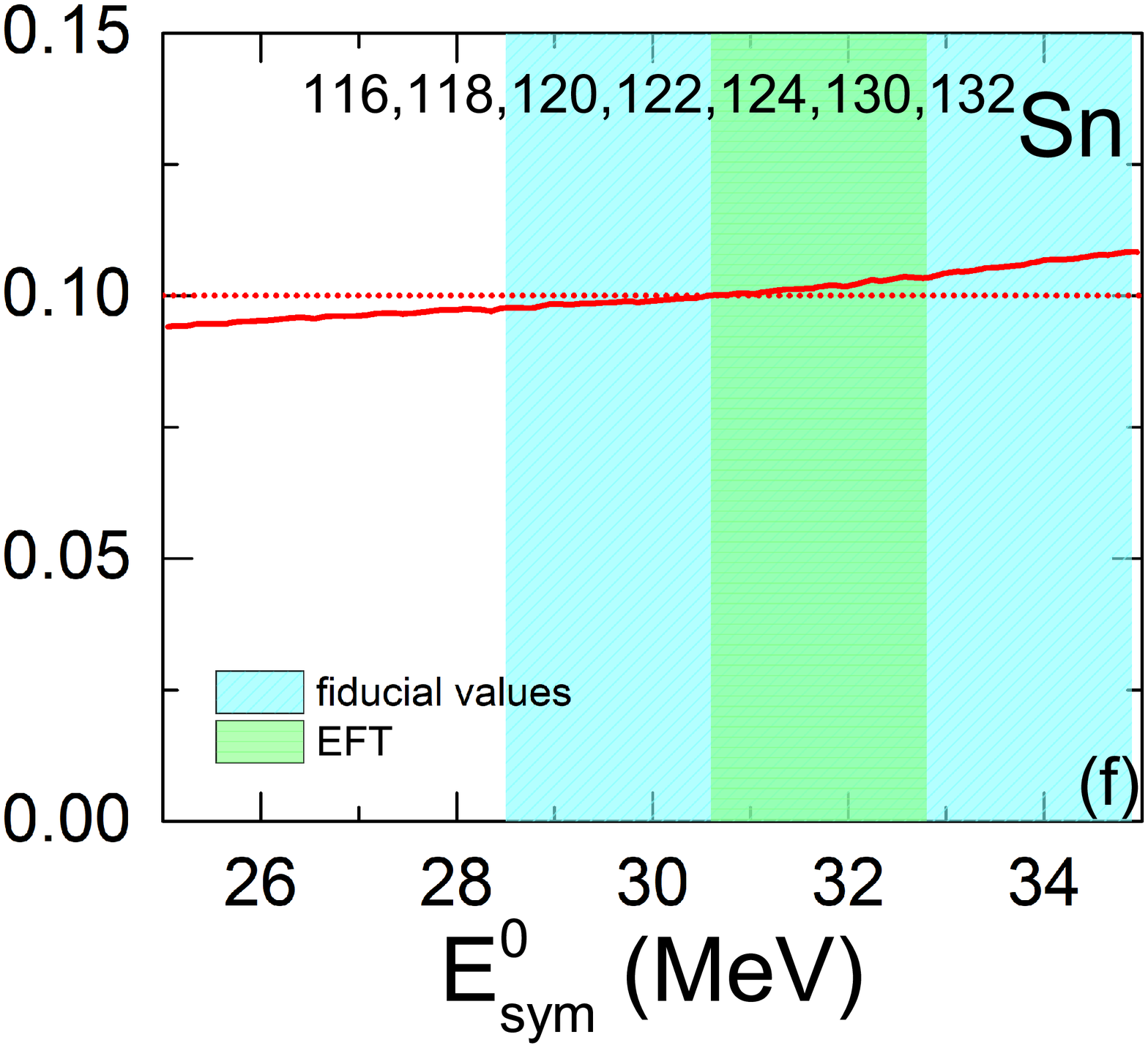}
	\caption{(Color online) Upper: Correlated posterior PDFs from neutron skin thicknesses in Sn isotopes in the $L-m_v^\star/m$ plane (a), the $E_{sym}^0-m_v^\star/m$ plane (b), and the $E_{sym}^0-L$ plane (c); Lower: Prior (dotted lines) and posterior (solid lines) probability distributions of $m_v^\star/m$ (d), $L$ (e), and $E_{sym}^0$ (f), with the band of fiducial values~\cite{BAL13,Oer17}, the $68\%$ confidence band from the EFT analysis~\cite{Ohio20}, and the PDF from the neutron star analysis~\cite{Xie19} (dashed line) plotted for comparisons.} \label{fig1}
\end{figure*}

\begin{table}\small
  \caption{The slope parameter $L$ of the symmetry energy at $68\%$ confidence level from real and imagined neutron skin thickness data of various nuclei used in this study, with the mean values calculated according to Eq.~(\ref{mean}), the MAP values from fitting the peaks of the PDFs using a Gaussian function, and the confidence intervals obtained by using Eq.~(\ref{confidence}).}
    \begin{tabular}{|c|c|c|c|}
   \hline
   Nucleus & $\Delta r_{np}$ (fm)  & $L$(mean) (MeV) & $L$(MAP) (MeV) \\
   \hline
   \hline
    $^{116}$Sn & $0.110 \pm 0.018$ &  &\\
    $^{118}$Sn & $0.145 \pm 0.016$ &  &\\
    $^{120}$Sn & $0.147 \pm 0.033$ &  &\\
    $^{122}$Sn & $0.146 \pm 0.016$ & $45.5^{+26.5}_{-21.6}$ & $53.4^{+18.6}_{-29.5}$\\
    $^{124}$Sn & $0.185 \pm 0.017$ &  &\\
    $^{130}$Sn & $0.23 \pm 0.04$ &  &\\
    $^{132}$Sn & $0.24 \pm 0.04$ &  &\\
   \hline
   \hline
    $^{208}$Pb & $0.15 \pm 0.02$ &  $35.6^{+19.1}_{-25.8}$ & $35.5^{+19.2}_{-25.7}$ \\
   \hline
    $^{208}$Pb & $0.15 \pm 0.04$ &  $42.0^{+18.4}_{-35.2}$ & $36.3^{+24.1}_{-29.5}$ \\
   \hline
    $^{208}$Pb & $0.15 \pm 0.06$ &  $48.1^{+16.9}_{-45.6}$ & $36.3^{+28.7}_{-33.8}$ \\
   \hline
    $^{208}$Pb & $0.20 \pm 0.02$ &  $65.2^{+26.4}_{-15.6}$ & $74.8^{+16.8}_{-25.2}$ \\
   \hline
    $^{208}$Pb & $0.20 \pm 0.04$ &  $64.2^{+36.6}_{-23.7}$ & $75.0^{+25.8}_{-34.5}$ \\
   \hline
    $^{208}$Pb & $0.20 \pm 0.06$ &  $63.2^{+41.2}_{-27.9}$ & $75.0^{+29.4}_{-39.7}$ \\
   \hline
    $^{208}$Pb & $0.30 \pm 0.02$ &  $112.5^{+7.5}_{-1.2}$ & $120.0^{+0.0}_{-8.7}$ \\
   \hline
    $^{208}$Pb & $0.30 \pm 0.04$ &  $102.5^{+17.5}_{-2.9}$ & $120.0^{+0.0}_{-20.4}$ \\
   \hline
    $^{208}$Pb & $0.30 \pm 0.06$ &  $91.0^{+29.0}_{-5.9}$ & $120.0^{+0.0}_{-34.9}$ \\
   \hline
   \hline
    $^{48}$Ca & $0.12 \pm 0.01$ & $14.4^{+3.8}_{-14.4}$ & $0.0^{+18.2}_{-0.0}$ \\
   \hline
    $^{48}$Ca & $0.12 \pm 0.02$ & $23.1^{+6.0}_{-23.1}$ & $0.0^{+29.1}_{-0.0}$ \\
   \hline
    $^{48}$Ca & $0.15 \pm 0.01$ & $30.8^{+9.8}_{-30.8}$ & $16.3^{+24.3}_{-16.3}$ \\
   \hline
    $^{48}$Ca & $0.15 \pm 0.02$ & $37.1^{+10.8}_{-37.1}$ & $20.8^{+27.1}_{-20.8}$ \\
   \hline
    $^{48}$Ca & $0.25 \pm 0.01$ & $114.3^{+5.7}_{-0.9}$ & $120.0^{+0.0}_{-6.6}$ \\
   \hline
    $^{48}$Ca & $0.25 \pm 0.02$ & $106.0^{+14.0}_{-2.0}$ & $120.0^{+0.0}_{-16.0}$ \\
   \hline
   \hline
    $^{208}$Pb and & $0.15 \pm 0.02$ &  &\\
    Sn isotopes & Refs.~\cite{Ter08,Kli07} &   $38.1^{+23.7}_{-21.1}$  & $42.5^{+19.3}_{-25.5}$ \\
   \hline
    $^{208}$Pb and & $0.15 \pm 0.04$ &  &\\
    Sn isotopes & Refs.~\cite{Ter08,Kli07} &   $42.3^{+25.2}_{-21.5}$  & $49.3^{+18.2}_{-28.5}$ \\
   \hline
    $^{208}$Pb and & $0.15 \pm 0.06$ &  &\\
    Sn isotopes & Refs.~\cite{Ter08,Kli07} &   $43.9^{+25.9}_{-21.5}$  & $51.0^{+18.8}_{-28.6}$ \\
   \hline
    $^{208}$Pb and & $0.20 \pm 0.02$ &  &\\
    Sn isotopes & Refs.~\cite{Ter08,Kli07} &   $55.2^{+24.8}_{-15.7}$  & $64.0^{+16.0}_{-24.5}$ \\
   \hline
    $^{208}$Pb and & $0.20 \pm 0.04$ &  &\\
    Sn isotopes & Refs.~\cite{Ter08,Kli07} &   $49.2^{+26.8}_{-19.0}$  & $58.0^{+18.0}_{-27.8}$ \\
   \hline
    $^{208}$Pb and & $0.20 \pm 0.06$ &  &\\
    Sn isotopes & Refs.~\cite{Ter08,Kli07} &   $47.3^{+26.9}_{-20.2}$  & $56.1^{+18.1}_{-29.0}$ \\
   \hline
   \hline
    $^{208}$Pb & $0.15 \pm 0.02$ &  &\\
    and $^{48}$Ca & $0.12 \pm 0.01$ &  $24.8^{+13.9}_{-16.7}$  & $26.3^{+12.4}_{-18.2}$ \\
   \hline
    $^{208}$Pb & $0.15 \pm 0.02$ &  &\\
    and $^{48}$Ca & $0.15 \pm 0.01$ & $32.6^{+19.3}_{-22.8}$  & $40.1^{+11.8}_{-30.3}$ \\
   \hline
    $^{208}$Pb & $0.20 \pm 0.02$ &  &\\
    and $^{48}$Ca & $0.12 \pm 0.01$ & $37.1^{+22.0}_{-14.5}$  & $44.9^{+14.2}_{-22.3}$ \\
   \hline
    $^{208}$Pb & $0.20 \pm 0.02$ &  &\\
    and $^{48}$Ca & $0.15 \pm 0.01$ & $47.9^{+26.8}_{-16.0}$  & $61.5^{+13.2}_{-29.6}$ \\
   \hline
    \end{tabular}
  \label{T1}
\end{table}

After integrating one of the isovector model parameters $L$, $m_v^\star/m$, or $E_{sym}^0$ according to Eq.~(\ref{2dpdf}), the resulting correlated PDFs of the other two parameters are shown in the upper panels of Fig.~\ref{fig1}. It is seen that the $L$ parameter is strongly correlated with the isovector effective mass $m_v^\star/m$, with the latter weakly correlated with the $E_{sym}^0$ within their prior ranges considered, due to the decompositions of the $L$ and $E_{sym}^0$ parameters \cite{Xu10} according to the Hugenholtz-Van Hove theorem \cite{HVH}, see, the extensive review in Ref. \cite{PPNP-Li}. The anti-correlated PDF in the $L-E_{sym}^0$ plane is similar to the $L-E_{sym}^0$ correlation observed in Fig. 6 of Ref.~\cite{MSL0}, where the traditional $\chi^2$ fit was performed using the same MSL0 interaction within SHF to the empirical properties of nuclear matter and some properties of finite nuclei as well as the same set of the neutron skin thickness data of Sn isotopes.
This consistency is what one expects. However, the Bayesian analysis can go beyond what the traditional analysis can provide. The posterior PDFs of each model parameter after integrating all the others according to Eq.~(\ref{1dpdf}) are shown in the lower panels of Fig.~\ref{fig1}, where the ranges of $L$ and $E_{sym}^0$ from fiducial values~\cite{BAL13,Oer17}, from the EFT analysis~\cite{Ohio20}, and from the neutron star analysis~\cite{Xie19} are also plotted for comparisons. It is seen that with the neutron skin thickness data of Sn isotopes, the uniform prior distribution of $L$ within $(0,120)$ MeV changes to a posterior distribution peaking around 50 MeV, while those of $m^\star_v/m$ and $E_{sym}^0$ are not improved by much compared to their prior PDFs. More quantitatively, the $L$ is determined to be within $(23.9,72.0)$ MeV around the mean value $45.5$ MeV or the MAP value 53.4 MeV at the $68\%$ confidence level by the $\Delta r_{np}$ data of Sn isotopes. Due to the asymmetric PDF of $L$, the mean value is smaller than the MAP value, with the latter consistent with the fiducial value $L = 58.7 \pm 28.1$~MeV~\cite{BAL13,Oer17}, the $L = 59.8 \pm 4.1$~MeV from the EFT analysis~\cite{Ohio20}, and the $L=66_{-20}^{+12}$ MeV from the neutron star analysis~\cite{Xie19}.

\begin{figure*}[ht]
\includegraphics[scale=0.26]{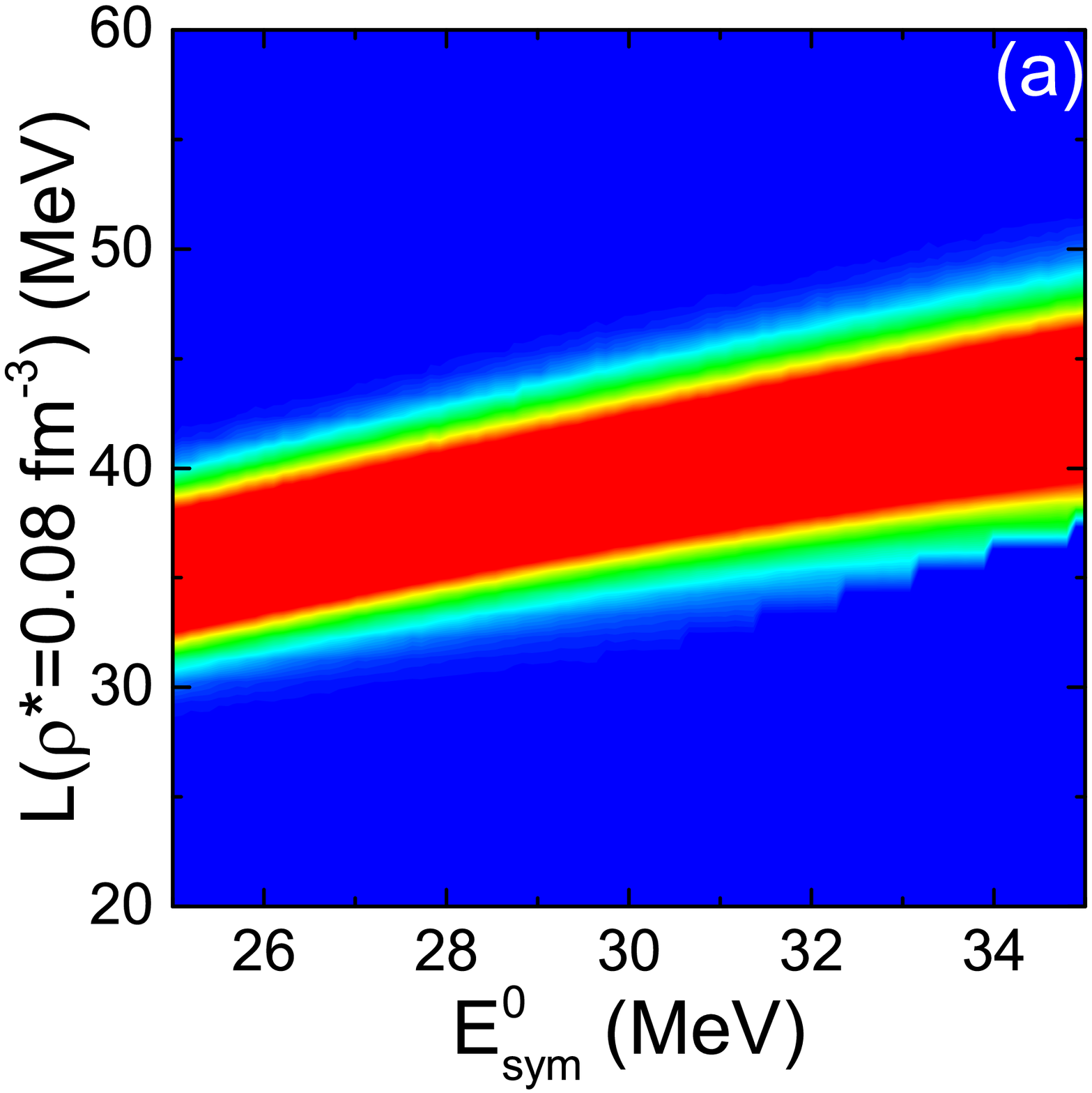}
\includegraphics[scale=0.26]{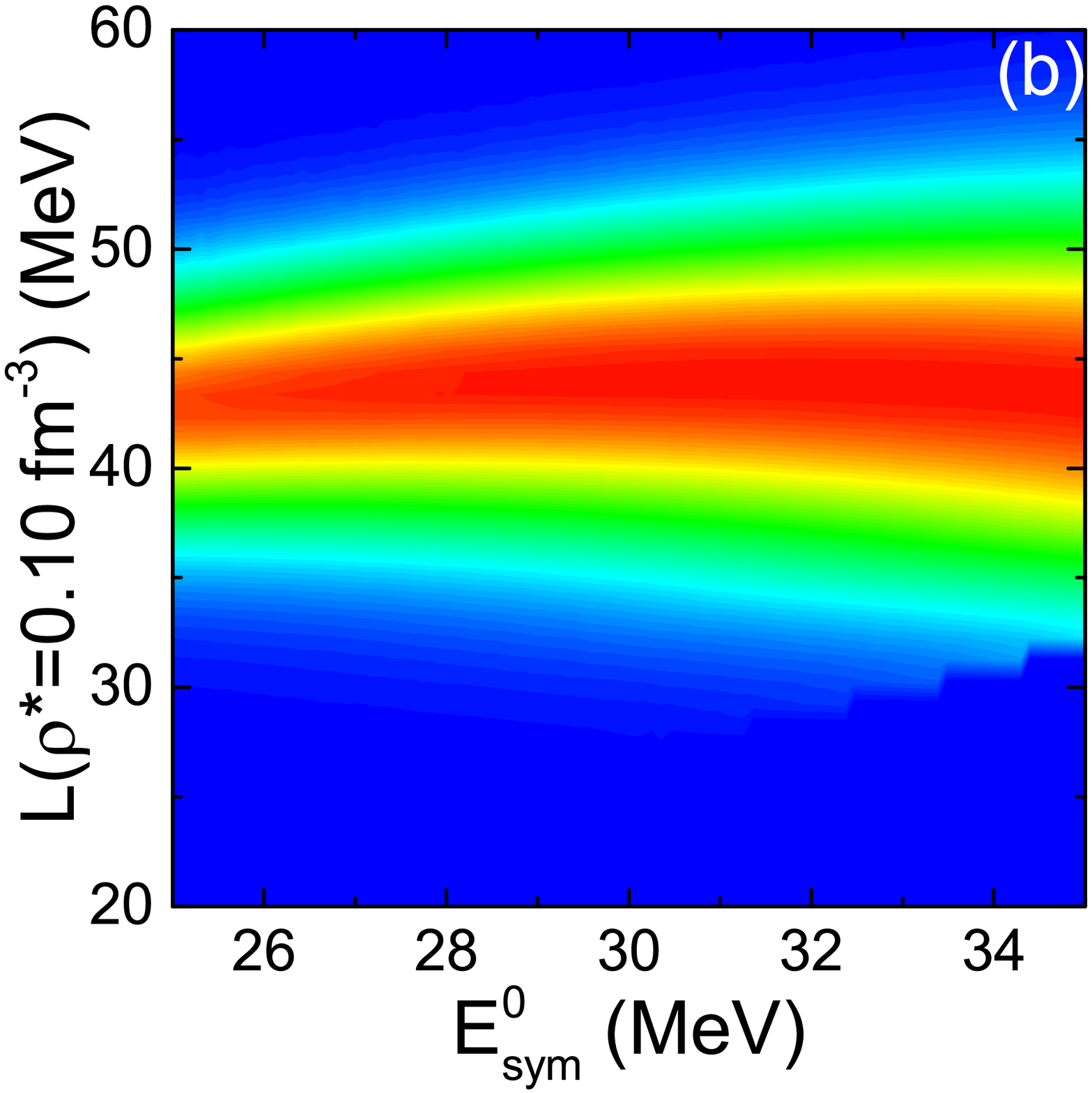}
\includegraphics[scale=0.26]{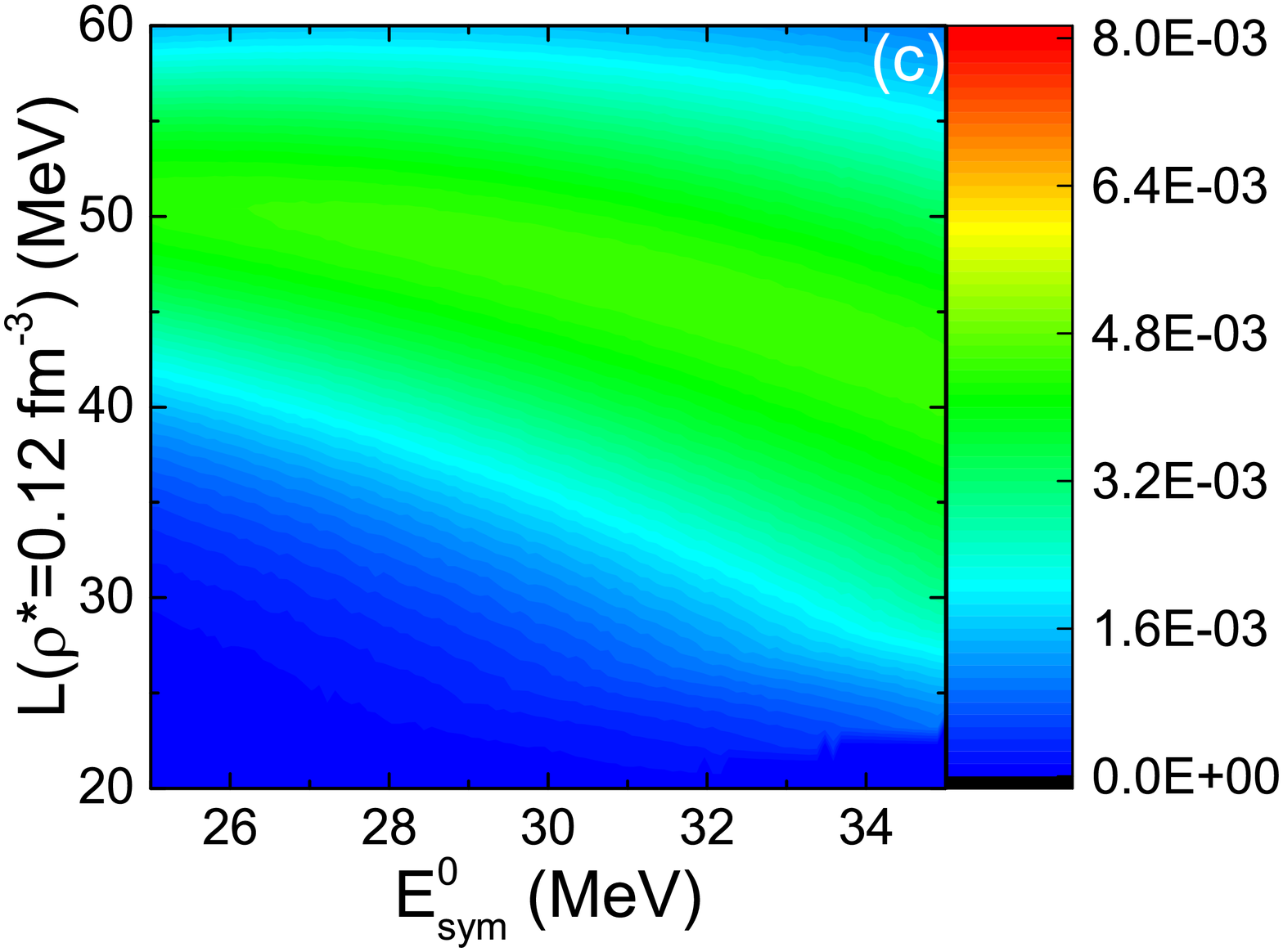}\\
\includegraphics[scale=0.24]{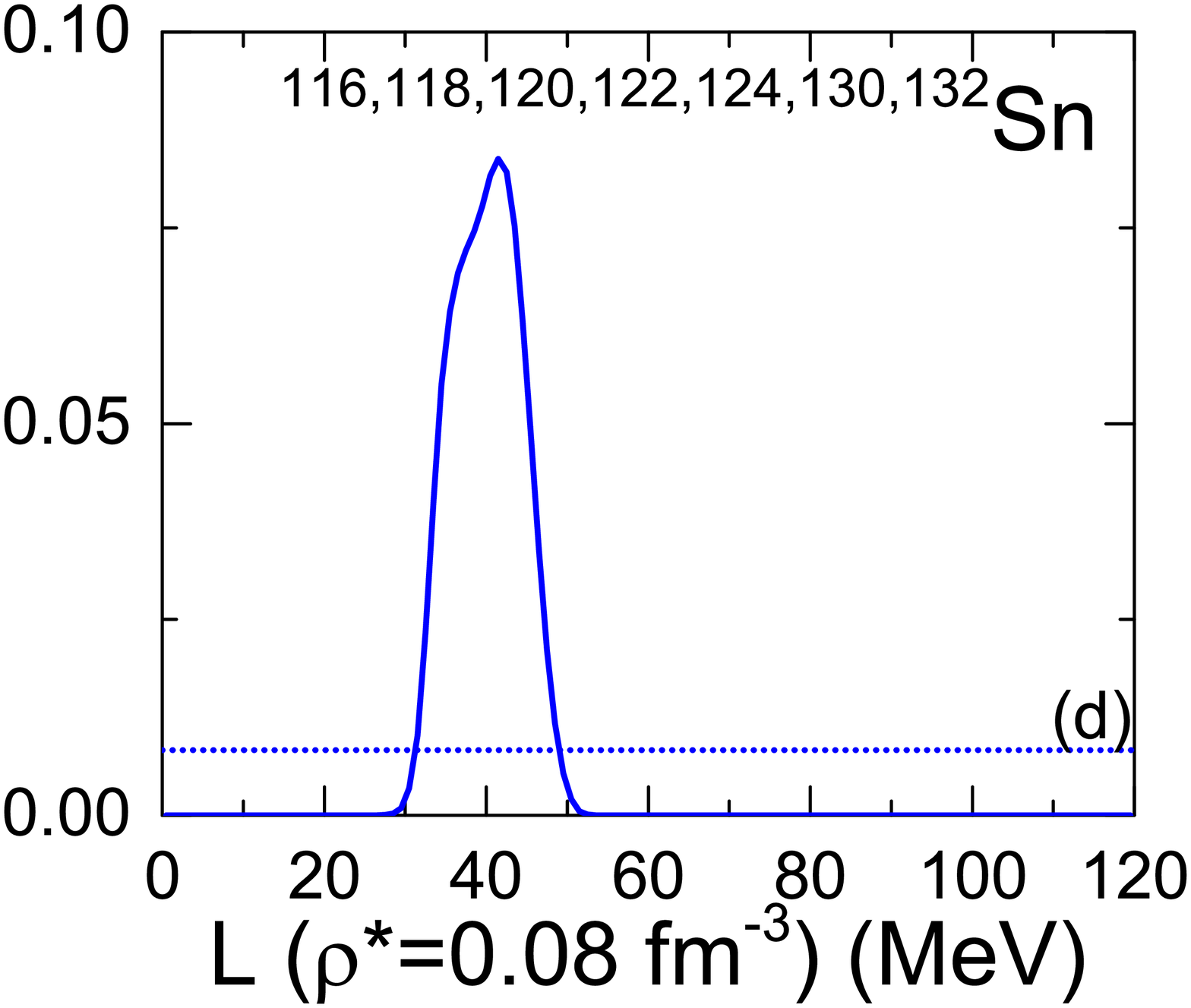}
\includegraphics[scale=0.24]{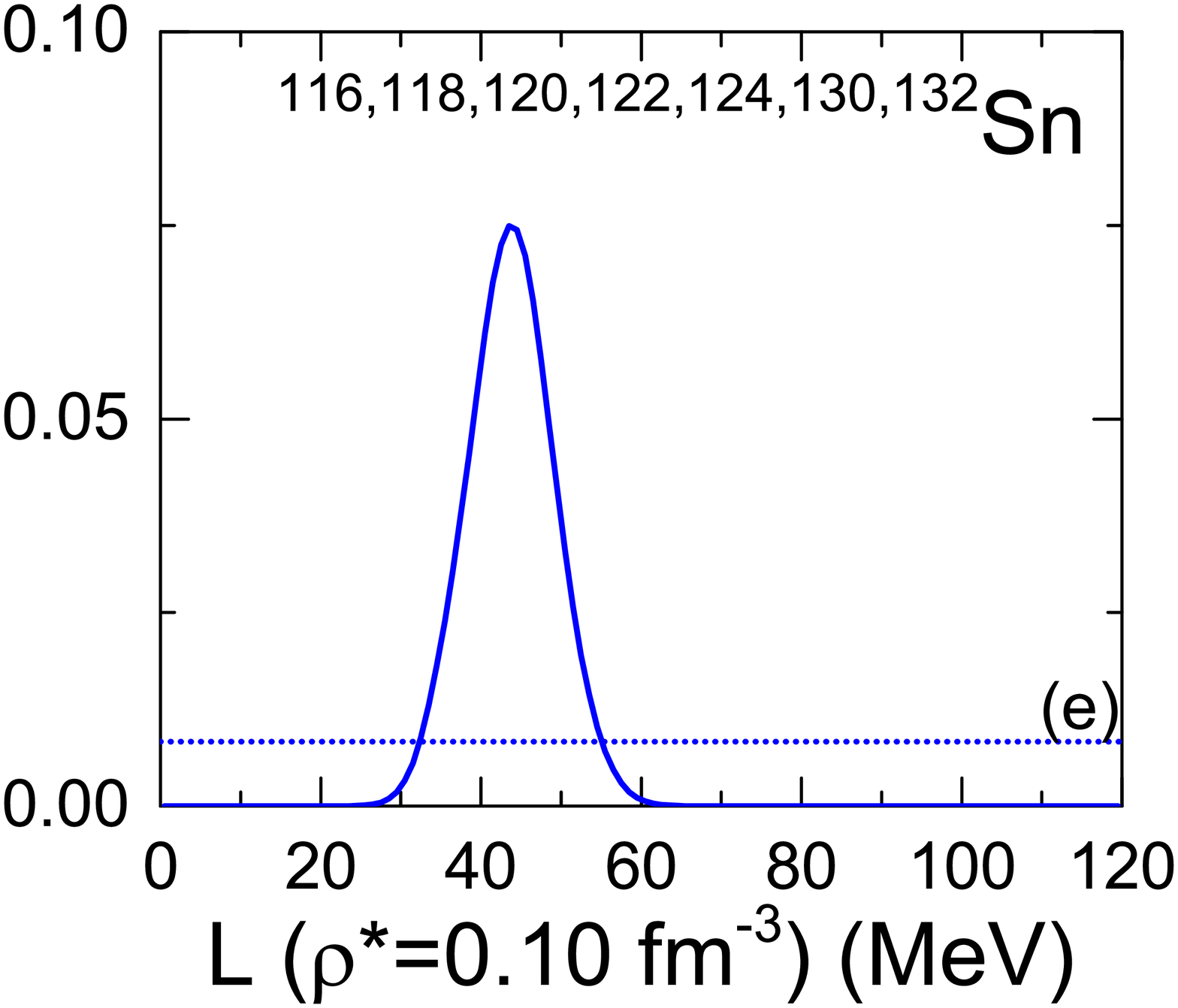}
\includegraphics[scale=0.24]{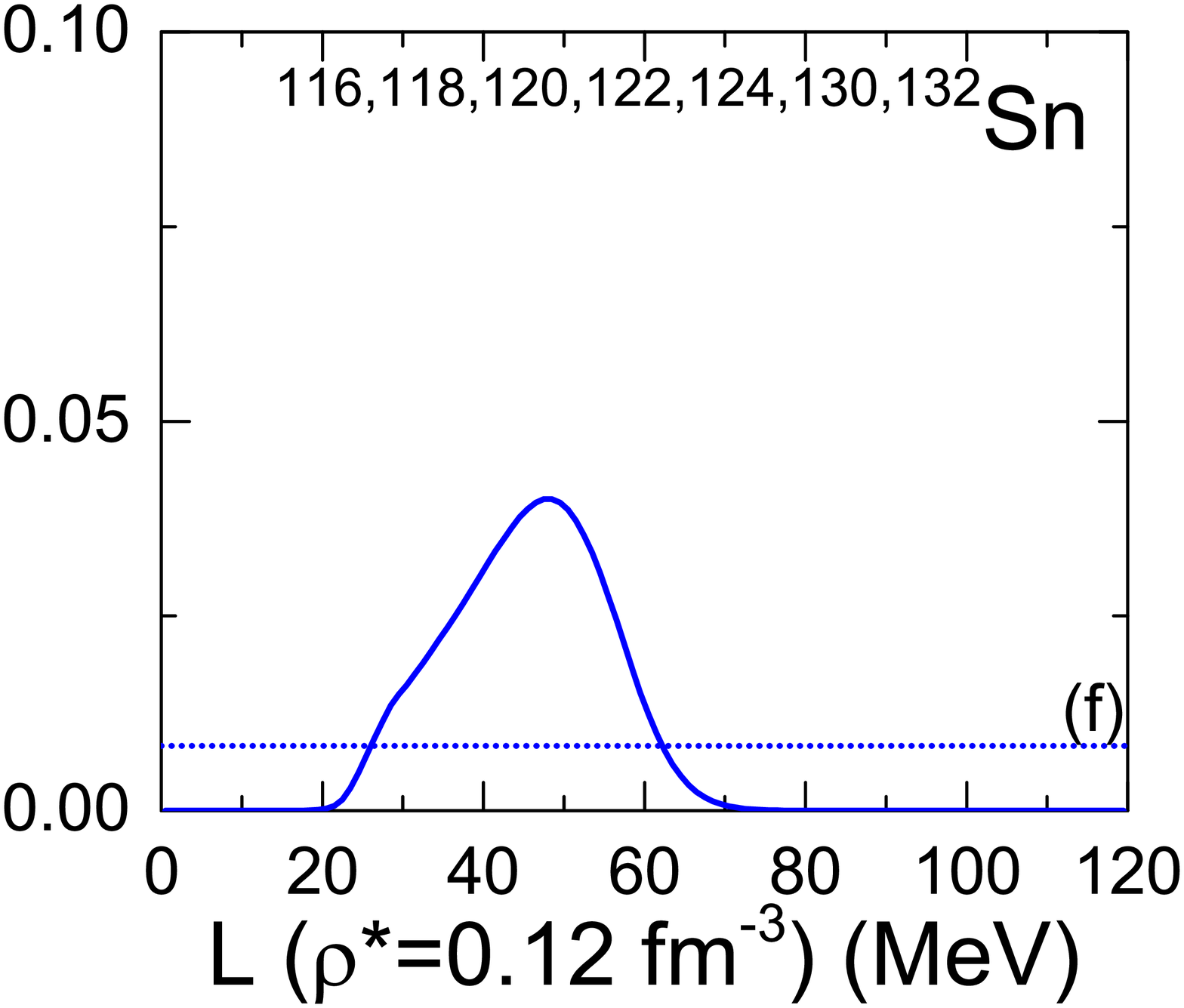}
	\caption{(Color online) Upper: Correlated posterior PDFs from neutron skin thicknesses in Sn isotopes in the $E_{sym}^0-L(\rho^\star)$ plane for $\rho^\star=0.08$, 0.10, and 0.12 fm$^{-3}$; Lower: Prior (dotted lines) and posterior (solid lines) probability distributions of $L(\rho^\star)$.} \label{fig2}
\end{figure*}

The anti-correlation between the $E_{sym}^0$ and $L$ in Fig.~\ref{fig1}(c) deserves some discussions. As noticed before \cite{MSL0,Xu10,Zha13,Lat14}, this correlation is opposite to the positive correlation from studying nuclear giant resonances, heavy-ion collisions, and the electric dipole polarizability~\cite{Tri08,Lat14,Xu20}. The overlapping area of these opposite correlations played a critical role in finding the common constraints on the $E_{sym}^0-L$ plane \cite{Xu10,Li13,Zha13,Lat14}. However, its origin needs further understanding. For this purpose, shown in the upper panels of Fig.~\ref{fig2} are the correlated PDFs between the $E_{sym}^0$ and $L(\rho^\star)=3\rho^\star (dE_{sym}/d\rho)_{\rho^\star}$ at different subsaturation densities $\rho^\star$ using the same Bayesian analysis method. Here the $L(\rho^\star)$ calculated at $\rho^\star$ from the same density dependence of $E_{sym}(\rho)$ depends on the SHF parameters in the same way as the $E_{sym}^0$ and $L$ at $\rho_0$.  They are thus all correlated. It is interesting to see that at the density $\rho^\star$ smaller (larger) than 0.10 fm$^{-3}$ the $L(\rho^\star)$ and $E_{sym}^0$ are positively correlated (anti-correlated), while at $\rho^\star=0.10$ fm$^{-3}$ the PDF of $L(\rho^\star)$ is approximately independent of $E_{sym}^0$. Moreover, it is seen that at the $68\%$ confidence level, $L(\rho^\star=0.10~\text{fm}^{-3})$ is tightly constrained to $43.7^{+5.3}_{-5.3}$ MeV with an symmetric distribution, while the PDFs of $L(\rho^\star)$ are generally broader especially at higher densities. This shows that the neutron skin thicknesses in Sn isotopes determines most tightly the value of $L(\rho^\star)$ around $\rho^\star=0.10$ fm$^{-3}$, which is approximated the average density of a nucleus. We note that this finding is robust for different nuclei, since the neutron skin thickness in $^{208}$Pb and $^{48}$Ca is also found to be mostly determined by $L(\rho^\star=0.10~\text{fm}^{-3})$ as well, and this is consistent with that observed in Ref.~\cite{Zha13} within the traditional approach.

So, why are the $E_{sym}^0$ and $L$ at $\rho_0$ anti-correlated?  As shown in the upper panels of Fig.~\ref{fig2}, there is a clear tendency that their correlation changes from positive to negative as the density increases towards $\rho_0$. One can understand these numerical results by analytically investigating how the $E_{sym}^0$ and $L$ at $\rho_0$ are correlated when a constraint is applied to the function $E_{sym}(\rho)$ at a subsaturation density $\rho^*$. In the Appendix A, using a general form of the symmetry energy $E_{sym}(\rho) = E_{sym}^0 \cdot\left( \frac{\rho}{\rho_0}\right)^\gamma$ describing those predicted by SHF very well, we have shown analytically that the $E_{sym}^0$ and $L$ at $\rho_0$ are positively correlated if the observable used constrains the magnitude of $E_{sym}(\rho^*) $,
while a negative correlation appears if the observable constrains the $L(\rho^*)$ at $\rho^*$. In the situation here, the neutron skin thickness constrained the $L(\rho^*)$ but not $E_{sym}(\rho^*)$ around $\rho^\star=0.10$ fm$^{-3}$. Consequently, the neutron skin constraint leads to a negative correlation between the $E_{sym}^0$ and $L$ at $\rho_0$. We have also noticed that the strength of the anti-correlation between $L(\rho^\star)$ $\rho^\star=0.12$ fm$^{-3}$ and $E_{sym}^0$ is much weaker compared with that at smaller $\rho^\star$. This indicates the difficulty of constraining the symmetry energy at the saturation density using the neutron skin data. Basically, the latter determines the slope of $E_{sym}$ at 0.10 fm$^{-3}$, while the information about the $E_{sym}$ at higher densities is from extrapolating the underlying energy density functional. Thus, while the neutron skin data may be model independent and very precise, the extraction of $E_{sym}$ or $L$ at $\rho_0$ from the neutron skin data is also model dependent. The correlation between the neutron-skin thickness of nuclei and the radii of neutron stars is even weaker and very model dependent as demonstrated numerically already in Ref.~\cite{Burg}. Here we used the SHF functional in our Bayesian analysis, and it would be interesting to study in the future with other models.

It is interesting to note that in a recent Bayesian analysis \cite{Xu20} using the centroid energy $E_{-1}$ of the isovector giant dipole resonance in $^{208}$Pb as well as its electric polarizability $\alpha_D$,
it was found that these data determine the nuclear symmetry energy $E_{sym}$ at about $\rho^\star=0.05$ fm$^{-3}$ and the isovector nucleon effective mass $m_v^\star$ at $\rho_0$. At $90\%$ confidence level, $E_{sym}(\rho^\star) = 16.4 ^{+1.0} _{-0.9}$ MeV and $m_v^\star/m = 0.79 ^{+0.06} _{-0.06}$ were obtained around their mean values. Compared to what we have learned from the Bayesian analysis of neutron skin thicknesses of Sn isotopes, the results are complimentary for mapping out the density dependence of nuclear symmetry energy while their difference is completely understandable. Specifically, the neutron skin thickness is mostly dominated by the neutron pressure related to $L$~\cite{Bro00,Fur02}, while the giant resonances are affected by both the restoring force from the EOS and the nucleon effective mass~\cite{Zha16,Kon17,Xu20}.

\begin{figure}[ht]
	\includegraphics[scale=0.28]{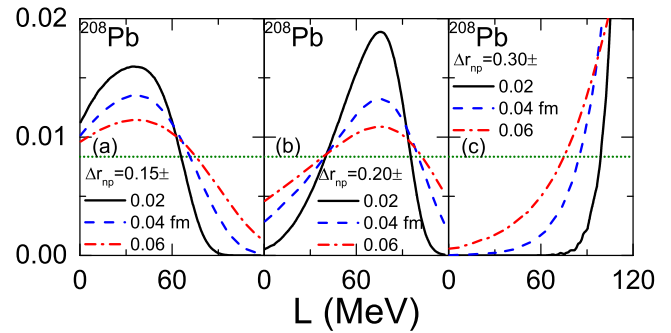}
	\caption{(Color online) Prior (dotted lines) and posterior probability distributions of $L$ from the imagined neutron skin thickness 0.15 (a), 0.20 (b), and 0.30 (c) fm in $^{208}$Pb with different error bars.} \label{fig3}
\end{figure}

\subsection{$L$ from imagined neutron skin thickness in $^{208}$Pb}

As discussed in the introduction, we want to know if and how new measurements can improve our knowledge about the symmetry energy, especially its slope parameter $L$ at $\rho_0$
with respect to what we learned from analyzing the Sn isotopes and neutron star data. Since the neutron skin thickness in $^{208}$Pb is still not well determined,
to illustrate how the uncertainties of $\Delta r_{np}$ in $^{208}$Pb may affect the extraction of $L$, we display in Fig.~\ref{fig3} its posterior PDFs by using the imagined neutron skin thickness data of $\Delta r_{np}=0.15$, 0.20, and 0.30 fm with different error bars. As one expects, a larger neutron skin thickness generally leads to a larger value of $L$. With $\Delta r_{np}=0.30$ fm, $L$ would peak outside the prior range of $(0,120)$ MeV if we enlarge it further, contradictory to most of the existing constraints listed in Refs.~\cite{BAL13,Oer17}. Of course, the PDFs become broader with a larger experimental error bar, showing a reduced constraining power on the PDF of $L$. The width of the PDF actually depends on the relative error bar of the experimental data, i.e., a smaller width in PDF is obtained with a larger mean value of the experimental data for the same absolute $1\sigma$ error bar as one expects.
The $L$ values at $68\%$ confidence level around the mean values and the MAP values from the real and imagined neutron skin thickness data of various nuclei used in this study are compared in Table \ref{T1}.

\begin{figure}[ht]
	\includegraphics[scale=0.5]{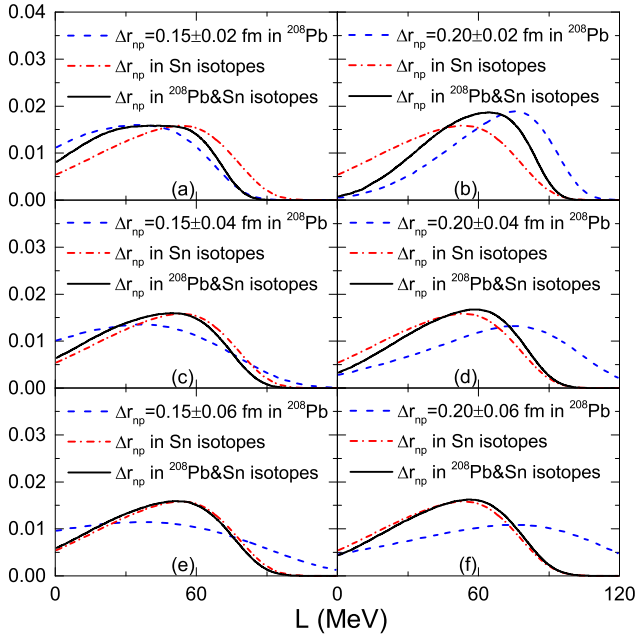}
	\caption{(Color online) Posterior probability distributions of $L$ from imagined neutron skin thicknesses with different mean values and error bars in $^{208}$Pb (dashed lines), from real neutron skin thickness data of Sn isotopes (dot-dashed lines), as well as from their combinations (solid lines).} \label{fig4}
\end{figure}

What further information on $L$ can the measurement of $\Delta r_{np}$ in $^{208}$Pb bring to us, in additional to our knowledge from analyzing the Sn isotopes? To answer this question, we compare in Fig.~\ref{fig4} the PDFs of $L$
from using the imagined $\Delta r_{np}=0.15$ (0.20) fm data of $^{208}$Pb with different error bars, the measured $\Delta r_{np}$ data of Sn isotopes, and the combined data, respectively. Since $\Delta r_{np}=0.15$ (0.20) fm of $^{208}$Pb leads to smaller (larger) $L$ values compared to that extracted from the $\Delta r_{np}$ data of Sn isotopes, the PDF of $L$ from the combined data is shifted and peaks at a smaller (larger) value. We found that a $\Delta r_{np}=0.17-0.18$ fm of $^{208}$Pb with an error bar of about 0.02 fm leads to a PDF of $L$ compatible with that from analyzing the Sn data. On the other hand, it is shown that larger error bars of $\Delta r_{np}$($^{208}$Pb) weaken the effects of incorporating the $^{208}$Pb data into the Bayesian analysis with the combined data. Quantitatively, an error bar as large as 0.06 fm for $\Delta r_{np}$($^{208}$Pb) leads to negligible improvements of the posterior PDF of $L$ extracted from the $\Delta r_{np}$ of Sn isotopes.

\subsection{$L$ from imagined neutron skin thickness in $^{48}$Ca}

\begin{figure}[ht]
	\includegraphics[scale=0.3]{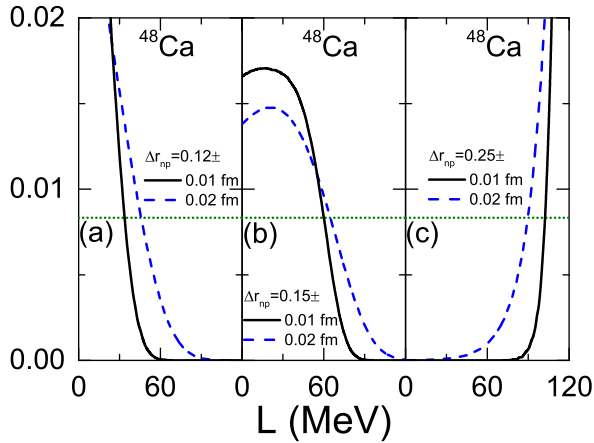}
	\caption{(Color online) Prior (dotted lines) and posterior probability distributions of $L$ from imagined neutron skin thicknesses 0.12 (a), 0.15 (b), and 0.25 (c) fm in $^{48}$Ca with different error bars.} \label{fig5}
\end{figure}

An {\it ab initio} calculation in Ref.~\cite{Hag16} has predicted that the neutron skin thickness in $^{48}$Ca is about $0.12-0.15$ fm, while it is predicted to be about 0.25 fm from a nonlocal dispersive optical-model analysis~\cite{Dickhoff}. Accordingly, here we consider three cases of $\Delta r_{np}=0.12$, $0.15$, and 0.25 fm with an $1\sigma$ error bar of 0.01 and 0.02 fm, respectively. The resulting PDFs of $L$ are displayed in Fig.~\ref{fig5} with the prior range of $(0,120)$ MeV. With $\Delta r_{np}=0.12$ or 0.25 fm, the posterior PDF of $L$ would peak out of the prior range of $(0,120)$ MeV if allowed, incompatible with
the known range of $L$ from earlier analyses~\cite{BAL13,Oer17}.
\begin{figure}[ht]
	\includegraphics[scale=0.35]{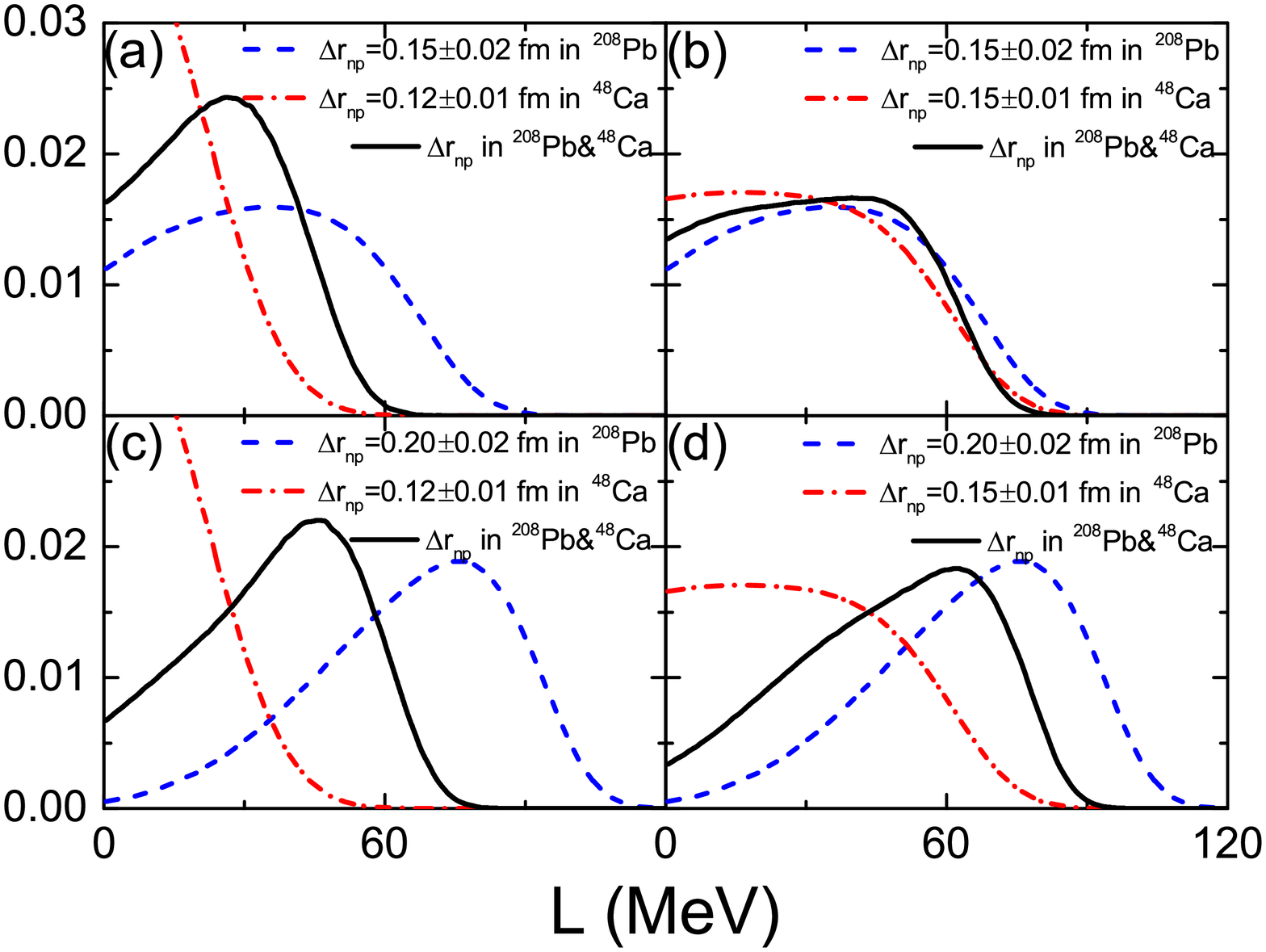}
	\caption{(Color online) Posterior probability distributions of $L$ from imagined neutron skin thicknesses in $^{208}$Pb (dashed lines), imagined neutron skin thicknesses in $^{48}$Ca (dot-dashed lines), as well as from their different combinations (solid lines).} \label{fig6}
\end{figure}

To compare the posterior PDFs of $L$ from analyzing the $\Delta r_{np}$ in $^{48}$Ca with those in the case of $^{208}$Pb, one needs to compare results with approximately the same relative values of both the neutron skin thicknesses and their error bars with respect to the radii of the two nuclei. Comparing the fractions $\Delta r_{np}/R$ of $^{48}$Ca and $^{208}$Pb, the radius $R$ of $^{48}$Ca is about 4.3 fm, while that of $^{208}$Pb is about 7.0 fm. For the same $\Delta r_{np}=0.15$ fm, it is about $3.5\%$ of the radius for $^{48}$Ca but only $2\%$ for $^{208}$Pb.
For the same reason, with the same imagined $0.02$ fm absolute error bar the relative error for the neutron skin thickness is actually larger for $^{48}$Ca than for $^{208}$Pb.

What further information on $L$ can the new neutron skin thickness measurement of $^{48}$Ca bring to us?
To answer this question, we have done Bayesian analyses by using both the imagined experimental data for $^{208}$Pb and $^{48}$Ca. The resulting posterior PDFs of $L$ from different combinations of $\Delta r_{np}$ in $^{208}$Pb and $^{48}$Ca are shown in Fig.~\ref{fig6}. Due to the different constraints on $L$ from $\Delta r_{np}$ in $^{208}$Pb and $^{48}$Ca, it is seen that the posterior PDFs of $L$ indicated by the solid lines are in-between those from two separate analyses, with the dashed lines from only the $\Delta r_{np}$ in $^{208}$Pb and dot-dashed lines from only the $\Delta r_{np}$ in $^{48}$Ca, respectively. The corresponding $L$ values at 68\% confidence level around the mean values and the MAP values are listed in Table \ref{T1}. Again, the final PDFs also depend on the $1\sigma$ error bar of $\Delta r_{np}$. Using a larger $1\sigma$ error bar for the $\Delta r_{np}$ in $^{208}$Pb or $^{48}$Ca, the corresponding PDF of $L$ becomes broader and less important, and the posterior PDF of $L$ from the combined $\Delta r_{np}$ data is closer to the one with a smaller error bar. Our results indicate that it is better to analyze the $^{208}$Pb and $^{48}$Ca data separately, then compare the $L$ values extracted, instead of combining the data and extracting a common $L$. This is because the two nuclei have very different charge radii. Coulomb and other dynamical effects in the two nuclei may be very different unlike the neutron skins in isotope chains having the same charge.

\section{Summary and conclusions}
\label{summary}

In summary, within the Bayesian statistical framework using both real and imagined neutron skin thickness data in heavy and medium nuclei, we have investigated how the available and expected data may help improve our knowledge about
the density dependence of nuclear symmetry energy, especially its slope parameter $L$ at the saturation density of nuclear matter. Using the available data for Sn isotopes, we have not only extracted the posterior PDF of $L$ parameter as a useful reference for future studies with new data of high precisions from parity-violating electron scattering experiments, but also found the density region in which the neutron skin data is most sensitive to the variation of symmetry energy. We also demonstrated numerically and explained analytically why the magnitude and the slope parameter of symmetry energy at $\rho_0$ are anti-correlated when the experimental constraint on the neutron skin thickness is applied. Moreover, we compared the $L$ values extracted from the Bayesian analyses of the neutron skin data in Sn isotopes and observations of neutron stars. They are largely compatible within their 68\% confidence intervals.

Furthermore, we found that a neutron skin of the size $\Delta r_{np}=0.17-0.18$ fm in $^{208}$Pb with an  error bar of about 0.02 fm leads to a PDF of $L$ compatible with that from analyzing the Sn neutron skin data, while the $\Delta r_{np}(^{208}$Pb$)=0.30$ fm regardless of its error bar leads to a posterior PDF of $L$ largely incompatible with the results from analyzing neither the neutron star data nor the neutron skin data of Sn isotopes. In order to provide new information on $L$ compared to our current knowledge about it, the experimental error bar of $\Delta r_{np}$ in $^{208}$Pb should be at least smaller than 0.06 fm aimed by some current experiments. On the other hand, the $\Delta r_{np}(^{48}$Ca) needs to be larger than 0.15 fm but smaller than 0.25 fm for the extracted PDF of $L$ to be compatible with the Sn and/or neutron star results. To further improve our current knowledge about $L$ and distinguish its posterior PDFs in the examples considered in this work, better precisions of measurements leading to significantly less than $\pm 20$ MeV error bars for $L$ at 68\% confidence level are necessary.

\begin{acknowledgments}
JX acknowledges the National Natural Science Foundation of China under Grant No. 11922514. WJX acknowledges the National Natural Science Foundation of China under Grant No. 11505150. BAL acknowledges the U.S. Department of Energy, Office of Science, under Award Number DE-SC0013702, the CUSTIPEN (China-U.S. Theory Institute for Physics with Exotic Nuclei) under the US Department of Energy Grant No. DE-SC0009971.
\end{acknowledgments}

\begin{appendix}

\section{Intuitive discussions on the correlation between $L$ and $E_{sym}^0$
}
\label{app}

Here we discuss intuitively the correlation between the symmetry energy $E_{sym}^0$ at the saturation density and the slope parameter $L$ of the symmetry energy at the saturation density. We will show that their positive correlation means that the observable is dominated by the symmetry energy at a subsaturation density, while their negative correlation means that the observable is dominated by the slope parameter of the symmetry energy at a subsaturation density.


We illustrate the idea with a popularly used symmetry energy of the following form
\begin{equation}
E_{sym}(\rho) = E_{sym}^0 \cdot\left( \frac{\rho}{\rho_0}\right)^\gamma.
\end{equation}
Thus, the slope parameter $L$ of the symmetry energy can be expressed as
\begin{equation}
L = 3\rho_0 \left[ \frac{dE_{sym}(\rho)}{d\rho}\right]_{\rho_0}=3 E_{sym}^0 \gamma.
\end{equation}
For a fixed symmetry energy at a subsaturation density $\rho^\star$
\begin{equation}
E_{sym}(\rho^\star) = E_{sym}^0 \left( \frac{\rho^\star}{\rho_0}\right)^\gamma,
\end{equation}
the expression of $L$ in terms of $E_{sym}^0$ is
\begin{equation}
L = 3 E_{sym}(\rho^\star)\left[ \frac{E_{sym}^0}{E_{sym}(\rho^\star)}\right] \frac{\ln[E_{sym}^0/E_{sym}(\rho^\star)]}{\ln(\rho_0/\rho^\star)}.
\end{equation}
It is obviously seen that $L$ increases with increasing $E_{sym}^0$ (see Ref.~\cite{Xu20} as an example).
\end{appendix}
The slope parameter at $\rho^\star$ can be expressed as
\begin{equation}
L(\rho^\star) = 3\rho^\star \left[ \frac{dE_{sym}(\rho)}{d\rho}\right]_{\rho^\star}=L\left( \frac{\rho^\star}{\rho_0}\right)^\gamma,
\end{equation}
where $L(\rho^\star)$ is seen to be smaller than $L$. For a fixed $L(\rho^\star)$, the expression of $E_{sym}^0$ in terms of $L$ is
\begin{equation}
E_{sym}^0 = \frac{L(\rho^\star)}{3} \frac{\ln(\rho^\star/\rho_0)}{[L(\rho^\star)/L]\ln[L(\rho^\star)/L]}.
\end{equation}
The function $x\ln(x)$ is negative for $x<1$ and increases with increasing $x$ for $x>0.4$. Thus, $E_{sym}^0$ generally increases with increasing $x=L(\rho^\star)/L$. Since $L$ decreases with increasing $x$, this leads to an anti-correlation between $L$ and $E_{sym}^0$. This conclusion is general and helps us understand the results shown in Fig.~\ref{fig2} of the present manuscript.

\end{document}